\documentclass[12pt]{article}
\usepackage{float}
\usepackage{physics}
\usepackage{amsmath}
\usepackage{amssymb}
\usepackage{graphicx}
\usepackage{hyperref}
\usepackage[dvipsnames]{xcolor}
\usepackage{multirow}
\usepackage{multirow}
\usepackage{pdflscape}
\usepackage{appendix}
\usepackage{fmtcount} 
\usepackage{subcaption}
\usepackage{array,multirow}
\usepackage{tikz-feynman}
\usepackage{epsfig}
\usepackage{float}
\usepackage{geometry}
\usepackage{amsmath}
\usepackage{cite}
\usepackage{ctable}
\usepackage{enumitem}

\begin{document}
\title{Gauged $U(1)_{L_{\mu}-L_{\tau}}$ Symmetry and two-zero Textures of Inverse Neutrino Mass Matrix in light of Muon ($g-2$)}
\author{Labh Singh\thanks{sainilabh5@gmail.com}, Monal Kashav\thanks{monalkashav@gmail.com} and Surender Verma\thanks{s\_7verma@hpcu.ac.in, Corresponding Author}}
\date{%
Department of Physics and Astronomical Science\\
Central University of Himachal Pradesh\\
Dharamshala, India 176215
}

\maketitle

\begin{abstract}
 \noindent In the framework of anomaly free $U(1)_{L_{\mu}-L_{\tau}}$ model, charged scalar fields give rise to massive gauge boson ($Z_{\mu\tau}$) through spontaneous symmetry breaking. $Z_{\mu\tau}$ leads to one loop contribution to the muon anomalous magnetic moment. These scalar fields may, also, appear in the structure of right-handed neutrino mass matrix, thus, connecting the possible explanation of muon ($g-2$) and low energy neutrino phenomenology through $vevs$ associated with the scalar fields. In the present work, we consider textures of inverse neutrino mass matrix ($M_{\nu}^{-1}$) wherein any two elements of the mass matrix are zero. In this ansatz, with Dirac neutrino mass matrix diagonal, the zero(s) of right-handed Majorana neutrino mass matrix correspond to zero(s) in the low energy effective neutrino mass matrix (within Type-I seesaw). We have realized two such textures of $M_{\nu}^{-1}$ accommodating the muon ($g-2$) and low energy neutrino phenomenology. The requirement of successful explanation of muon ($g-2$), further, constrain the allowed parameter space of the model and results in sharp correlations amongst neutrino mixing angles, $CP$ invariants and effective Majorana mass ($M_{ee}$). The model explains muon ($g-2$) for $M_{Z_{\mu\tau}}$ in the range ($0.035$ GeV-$0.100$ GeV) and $g_{\mu\tau}\approx\mathcal{O}(10^{-4}$) which is found to be consistent with constraints coming from the experiments like CCFR, COHERENT, BABAR, NA62 and NA64.
 \vspace{1cm}\\
 \noindent\textbf{Keywords:} Muon $(g-2)$; Phenomenology; Neutrino mass matrix; Texture zeros.\\
\end{abstract}

\newpage
\section{Introduction}
\noindent The standard model (SM) of particle physics has been very successful in explaining interactions between fundamental particles and, at the same time, predicted a wide variety of phenomena. Despite immense success of the SM, it is facing a growing list of ``anomalies"- a significant experimental divergence from theoretical predictions. For example, unsolved problems like origin of light neutrino masses, matter-antimatter asymmetry, dark matter, muon anomalous magnetic moment etc. find no explanation within the SM. The experimental observations of sub-eV scale neutrino masses and large mixing in the leptonic sector provide cardinal evidences propounding physics beyond the standard model (BSM)\cite{neutrinodata}. The neutrino  flavor states ($\nu_{\alpha}(\alpha=e,\mu,\tau)$) are incoherent mixture of mass eigenstates ($\nu_{i}(i=1,2,3)$). The magnitude of mixing is parameterized in terms of three mixing angles ($\theta_{ij}(i,j=1,2,3;i<j)$) and one Dirac $CP$ phase ($\delta$). The Dirac $CP$ phase has not been observed experimentally, however, the recent measurements hint $\delta \approx -\pi/2$\cite{CPdata}. Additionally, two more $CP$ phases($\alpha,\beta$) appears for Majorana nature of neutrinos which have no influence on neutrino oscillations. Furthermore, the riddle of octant degeneracy ($\theta_{23}$ above or below 45$^\circ$) and mass ordering (normal($m_1<m_2<m_3$) or inverted hierarchy($m_3<m_1<m_2$)) still remains unresolved. Furthermore, although absolute scale of neutrino mass is unknown, we have upper bound on sum of neutrino mass, $\sum m_{i}< 0.12$ eV from  cosmological data\cite{Giusarma:2016phn,Aghanim:2018eyx}. In view of the above, the neutrino mass matrix, in general, contains more free parameters than one can measure experimentally so phenomenological ansatze are important to fully reconstruct the mass matrix in terms of less number of parameters to understand the underlying dynamics of neutrino mass generation.

\noindent The recent results from E989 experiment(Run I) \cite{g-2exp} at FermiLab for the precise measurement of muon anomalous magnetic moment  $a_{\mu}=(g-2)_{\mu}/2$, shows a discrepancy with the theoretical \cite{g-2th} prediction of the SM
\begin{equation}
a_{\mu}^{\text{FNAL}}=116592040(54)\cross 10^{-11},
\label{eqn1}
\end{equation}
\begin{equation}
a^{\text{SM}}_{\mu}=116591810(43)\cross 10^{-11},
\label{eqn2}
\end{equation}\\
which when amalgamated with the previous results of Brookhaven National Laboratory 
\begin{equation}
a_{\mu}^{\text{BNL}}=116592089(63)\cross 10^{-11},
\label{eqn3}
\end{equation}
raises the confidence level from 3.7$\sigma$ to 4.2$\sigma$ such that $\Delta a_{\mu}=a_{\mu}^{\text{EXP}}-a_{\mu}^{\text{SM}}=(251\pm 59)\cross 10^{-11}$, a compelling evidence of new physics.

\noindent The possible implication and interpretation of muon($g-2$) anomaly have been discussed in different frameworks such as 2HDM\cite{2HDM1,2HDM2,2HDM3,2HDM4,2HDM5},  model with axion-like particles (ALP)\cite{axiom}, $A_{4}$ modular symmetry\cite{modular}, vector-like leptons (VLL) \cite{VLL1,VLL2}, super-symmetric(SUSY) models\cite{SUSY1,SUSY2} $etc.$. In general, $U(1)_{L}$ and $U(1)_{B}$ are accidental symmetries in the SM leading to lepton and baryon number conservation, respectively, but are anomalous. However, the symmetries originating from the difference of any two charged lepton flavors, $i.e.$ $L_{\alpha}-L_{\beta}$, $(\alpha,\beta=e,\mu,\tau)$ are anomaly free. Among these $U(1)_{L_{\alpha}-L_{\beta}}$ symmetries, $U(1)_{L_{\mu}-L_{\tau}}$ gauge symmetry has been explored in different dimensions of neutrino mass model building scenarios\cite{model1,model2,model3}. $L_{\mu}-L_{\tau}$ extension of SM in the framework of  Type-I seesaw with one\cite{FIMP} and two complex scalar singlets\cite{Borah1} have been studied to explain the muon ($g-2$). In general, spontaneous symmetry breaking of $U(1)_{L_{\mu}-L_{\tau}}$ symmetry manifests the $L_{\mu}-L_{\tau}$ massive gauge boson ($Z_{\mu \tau}$). In the framework of  $U(1)_{L_{\mu}-L_{\tau}}$ symmetry, $Z_{\mu \tau}$ do not interact with electron and quarks, evading constraints from LEP\cite{LEP1,LEP2} and LHC\cite{LHC}. It interacts only with $\mu$ and $\tau$ flavors which may contribute significantly to muon magnetic moment. Also, $Z_{\mu\tau}$ gauge boson contributes to the muon neutrino trident (MNT) process which constrains the mass of new gauge boson $Z_{\mu\tau}\leq 300$ MeV for the explanation of muon ($g-2$)\cite{NTP}. The new developments in muon magnetic moment measurements compels to investigate theoretical models accommodating explanation of muon ($g-2$).

\noindent The phenomenological ansatze like texture zeroes\cite{texturezero1,texturezero2,texturezero3,texturezero4,texturezero5,texturezero6,texturezero7}, hybrid textures\cite{hybrid1,hybrid2,hybrid3,hybrid4}, magic symmetry\cite{magic1,magic2,magic3,magic4} lead to interesting predictions and correlation among low-energy observables. Also, texture zeroes in inverse neutrino mass matrices($M_{\nu}^{-1}$) are imperative in the sense, in diagonal charged lepton and Dirac mass basis, zeroes in right-handed neutrino mass matrix corresponds to zeroes in $M_{\nu}^{-1}$. The phenomenological implications of inverse neutrino mass textures have been studied in Refs.\cite{verma-inverse,lavoura,zhao}. Recently, the authors have investigated all possible two-zero texture inverse neutrino mass matrices ($M_{\nu}^{-1}$) in light of large mixing angle(LMA) and $dark$-large mixing angle(DLMA) solutions of neutrino mixing paradigm\cite{Singh}. Out of fifteen possible two-zero $M_{\nu}^{-1}$ textures only seven are found to be in consonance with current neutrino oscillation data. In the present work, we have realized two such textures $D_1$ and $E_{1}$( for notation of textures see Ref.\cite{Singh}) accommodating muon ($g-2$) anomaly and neutrino oscillation data, simultaneously. We have employed $U(1)_{L_{\mu}-L_{\tau}}$ symmetry extending the SM with three right-handed neutrinos and three scalar singlet fields. The gauge boson contributing to the possible explanation of muon anomalous magnetic moment, further, constrain the allowed parameter space of these textures.

The rest of the paper is organised as follows. In Section 2, we have discussed the $U(1)_{L_{\mu}-L_{\tau}}$ model and corresponding charge assignments resulting in two-zero $M_{\nu}^{-1}$. The details of numerical analysis and consequent discussion have been elaborated in Section 3. Finally, in Section 4, we summarize our conclusions.  

\section{$U(1)_{L_{\mu}-L_{\tau}}$ Model}
\noindent We have extended the SM field content with three heavy right-handed neutrinos ($N_{e}, N_{\mu}, N_{\tau}$) having $L_{\mu}-L_{\tau}$ charges ($0, 1, -1$), respectively, leading to Type-I seesaw origin of light neutrino masses. In the scalar sector, three singlet scalar fields $\Phi_{i}$ ($\textit{i}=1, 2, 3$) with non-zero $L_{\mu}-L_{\tau}$ charges have been employed. It is known that zeroes in $M_{R}$ are identical to zeroes in $M_{\nu}^{-1}$ if Dirac and charged lepton mass matrices ($M_D$ and $M_{\ell}$) are diagonal. Therefore, the charge assignments under $U(1)_{L_{\mu}-L_{\tau}}$ are chosen in such a way that $M_{\ell}$ and $M_{D}$ are diagonal, in the model. $\Phi_{i}$ breaks the $U(1)_{L_{\mu}-L_{\tau}}$ symmetry by acquiring vacuum expectation values ($vevs$) $v_{i}$ $(i=1,2,3)$ consequently giving mass to the new $U(1)_{L_{\mu}-L_{\tau}}$ gauge boson $Z_{\mu\tau}$. Also, the $Z_{4}$ symmetry have been used to constrain the structure of the Yukawa Lagrangian. The particle content with respective gauge charges under $SU(2)_{L}\cross U(1)_{Y}\cross U(1)_{L_{\mu}-L_{\tau}}\cross Z_{4}$ symmetry is shown in Table \ref{tab1}.
The Lagrangian for our model is given by
\begin{eqnarray}
\mathcal{L}=\mathcal{L}_{SM}+\mathcal{L}_{N}+\mathcal{L}_{gauge}+\mathcal{L}_{scalar},
\label{eqn4}
\end{eqnarray}
where $\mathcal{L}_{SM}$ is SM Lagrangian, $\mathcal{L}_{N}$ is Lagrangian for the right-handed neutrinos (RHN) which contains kinetic and mass terms. $\mathcal{L}_{gauge}$ includes gauge kinetic terms for new fields whereas $\mathcal{L}_{scalar}$ includes the form of scalar potential. It is to be noted that $\mathcal{L}_{SM}$ includes the charged lepton mass terms which is diagonal under the assignments in Table \ref{tab1}.\\
\noindent The new gauge kinetic terms that appears in the Lagrangian are
\begin{eqnarray}
\mathcal{L}_{gauge}=\dfrac{1}{4}\left( Z_{\mu\tau}\right)^{\gamma\eta}\left( Z_{\mu\tau}\right)_{\gamma\eta}-\dfrac{\epsilon}{2}\left( Z_{\mu\tau}\right)^{\gamma\eta}B_{\gamma\eta},
\label{eqn6}
\end{eqnarray}
where $\left( Z_{\mu\tau}\right)^{\gamma\eta}=\partial^{\gamma}\left( Z_{\mu\tau}\right)^{\eta}-\partial^{\eta}\left( Z_{\mu\tau}\right)^{\gamma}$ is the field strength tensor for new gauge boson, $Z_{\mu\tau}$, while the second term in the Eqn.(\ref{eqn6}) denotes the kinetic mixing of $U(1)_{Y}$ and  $U(1)_{L_{\mu}-L_{\tau}}$ gauge sectors, where $\epsilon$ is kinetic mixing parameter.\\
\noindent The Lagrangian of the scalar sector can be written as
\begin{eqnarray}
\mathcal{L}_{scalar}=(D_{\gamma}\Phi_{i})^{\dagger}(D^{\gamma}\Phi_{i})-V(H, \Phi_{i}),\hspace{0.3cm} \textit{i}=1, 2, 3.
\label{eqn8}
\end{eqnarray}
The covariant derivative $D_{\gamma}$ is defined as
\begin{eqnarray}
D_{\gamma}=\partial_{\gamma}-ig\dfrac{\tau}{2}.W_{\gamma}-ig^{'}\dfrac{Y}{2}B_{\gamma},
\label{eqn9}
\end{eqnarray}
where $g$ and $g^{'}$ are the coupling constants associated with $W_{\gamma}$ and $B_{\gamma}$ gauge fields, respectively.

The scalar potential is given by
\begin{eqnarray}
\begin{aligned}
V(H,\Phi_{i})=&-\mu_{\Phi_{i}}^{2}(\Phi_{i}^{\dagger}\Phi_{i})^{2}+\lambda_{\phi_{i}}(\Phi_{i}^{\dagger}\Phi_{i})^{2}+\lambda_{H\Phi_{i}}(H^{\dagger}H)(\Phi_{i}^{\dagger}\Phi_{i})+\lambda_{\Phi_{1} \Phi_{2}}(\Phi_{1}^{\dagger}\Phi_{1})(\Phi_{2}^{\dagger}\Phi_{2})\\
&+[\mu_{12}\Phi_{1}^{2}\Phi_{2}^{\dagger}+H.c.]+\lambda_{\Phi_{1} \Phi_{3}}(\Phi_{1}^{\dagger}\Phi_{1})(\Phi_{3}^{\dagger}\Phi_{3})+\lambda_{\Phi_{2} \Phi_{3}}(\Phi_{2}^{\dagger}\Phi_{2})(\Phi_{3}^{\dagger}\Phi_{3}).
\end{aligned}
\label{eqn10}
\end{eqnarray}
where $i=1, 2, 3$. The neutral component of SM Higgs($H$) breaks the electroweak symmetry spontaneously whereas singlets $\Phi_{1,2,3}$ breaks the $L_{\mu}-L_{\tau}$ gauge symmetry after acquiring the $vevs$ $v_{1,2,3}$. The Yukawa Lagrangian for charged leptons is given by
\begin{eqnarray}
\mathcal{L_{\ell}}=-Y_{\ell_{e}}\Bar{L}_{e}H e_{R}-Y_{\ell_{\mu}}\Bar{L}_{\mu}H \mu_{R}-Y_{\ell_{\tau}}\Bar{L}_{\tau} H \tau_{R}+h.c., 
\end{eqnarray} which leads to charged lepton mass matrix diagonal,  $M_{\ell}$=$\frac{v}{\sqrt{2}}$diag($Y_{\ell_{e}},Y_{\ell_{\mu}},Y_{\ell_{\tau}}$), where $Y_{\ell_{i}}$ with $i=e,\mu,\tau$ are the Yukawa couplings.
The Lagrangian relevant for neutrino mass is given by
\begin{eqnarray}
\begin{aligned}
\mathcal{L_{N}}=&\overline{N_{\mu}}i\gamma^{\mu}D_{\mu}N_{\mu}+\overline{N_{\tau}}i\gamma^{\tau}D_{\tau}N_{\tau}-\dfrac{1}{2}M N_{e}N_{e}-Y_{e\tau}\Phi_{1}N_{e}N_{\tau}-Y_{e\mu}\Phi_{3}N_{e}N_{\mu}\\
&-Y_{\tau \tau}\Phi_{2}N_{\tau}N_{\tau}-Y_{D_{e}}\Bar{L}_{e}\tilde{H}N_{e}-Y_{D_{\mu}}\Bar{L}_{\mu}\tilde{H}N_{\mu}-Y_{D_{\tau}}\Bar{L}_{\tau}\tilde{H}N_{\tau}+h.c.,
\end{aligned}
\label{eqn5}
\end{eqnarray}
where $\tilde{H}=i\sigma_{2}H^{*}$ and $M$ is a constant with dimension of mass.

\begin{table}[t]
    \centering
    \begin{tabular}{|c| c c c c c c c c c c c c c|}
    \hline
    Symmetry & $L_{e}$ &$L_{\mu}$&$L_{\tau}$& $e_R$ & $\mu_R$ & $\tau_R$ & $N_{e}$ & $N_{\mu}$ & $N_{\tau}$ & H & $\Phi_{1}$ & $\Phi_{2}$&$\Phi_{3}$\\
    \hline
    $SU(2)_L$ & 2 & 2 & 2 & 1 & 1 & 1 & 1 & 1 & 1 & 2 & 1 &1&1\\
    \hline
    $U(1)_{Y}$ & -$\frac{1}{2}$& -$\frac{1}{2}$ & -$\frac{1}{2}$ & -1 & -1 & -1 & 0 & 0 & 0 &$\frac{1}{2}$ & 0&0&0\\
     \hline
     $U(1)_{L_{\mu}-L_{\tau}}$ & 0& 1 & -1 & 0 & 1 & -1 & 0 & 1 & -1 & 0 & 1 &2&-1\\
     \hline
      $Z_{4}$ & -1& 1 & i & -1 & 1 & i & -1 & 1 & i & 1 & i &-1&-1\\
     \hline
    \end{tabular}
    \caption{The field content of the model with respective charge assignments under $SU(2)_{L}\cross U(1)_{Y}\cross U(1)_{L_{\mu}-L_{\tau}}\cross Z_{4}$.  }
    \label{tab1}
\end{table}

 After expanding the kinetic term in Eqn.(\ref{eqn8}), the mass of new  $U(1)_{L_{\mu}-L_{\tau}}$ gauge boson can found to be $M_{Z_{\mu\tau}}$=$g_{\mu\tau} \sqrt{v_1^{2}+4v_2^{2}+v_3^{2}}$, where $g_{\mu\tau}$ is the $L_{\mu}-L_{\tau}$ gauge coupling.
 
 Using Eqn.(\ref{eqn5}), the Dirac mass matrix is given by
\begin{eqnarray}
M_{D}=
\begin{pmatrix}
d_{e} &  0 & 0 \\
0 & d_{\mu}  & 0 \\
0 & 0  & d_{\tau}\\
\end{pmatrix},
\label{eqn12}
\end{eqnarray}

\noindent where $d_{\alpha}=\frac{Y_{D_{\alpha}} v}{\sqrt{2}}$ with $\alpha= e, \mu, \tau$. $Y_{D_{\alpha}}$ are real Yukawa couplings and $\frac{v}{\sqrt{2}}$ is the $vev$ of SM Higgs doublet, $H$. Using Eqn.(\ref{eqn5}), the right-handed Majorana mass matrix ($M_{R}$) is given by
\begin{eqnarray}
M_{R} = 
\begin{pmatrix}
M & Y_{e\mu} v_{3}  & Y_{e\tau} v_{1} e^{i\xi}\\
Y_{e\mu} v_{3} & 0  & 0 \\
Y_{e\tau} v_{1}e^{i\xi} & 0  & Y_{\tau \tau} v_{2}\\
\end{pmatrix},
\label{eqn11}
\end{eqnarray}
where, in general, the elements of $M_{R}$ are complex. By redefinition of the fields, $\xi$ is the only remaining irremovable phase.
Thus, $M_{R}$ depends on four real parameters $M$, $Y_{e\mu}$, $Y_{e\tau}$ and $Y_{\tau\tau}$ and a complex phase $\xi$. As a consequence of diagonal $M_D$ and $M_{\ell}$ the non-trivial neutrino mixing will arise from $M_{R}$.
\\
Within the paradigm of Type-I seesaw, the inverse neutrino mass matrix can be written as
\begin{eqnarray}\label{eqn13}
M_{\nu}^{-1}=-(M_{D}^{T})^{-1}M_{R}M_{D}^{-1}.
\end{eqnarray}
Using $M_{D}$ and $M_{R}$ given in Eqns.(\ref{eqn12}) and (\ref{eqn11}), respectively, alongwith Eqn. (\ref{eqn13}), $M_{\nu}^{-1}$ is given by
\begin{eqnarray}
M_{\nu}^{-1}=
\begin{pmatrix}
-\dfrac{M}{d_{e}^{2}} & \dfrac{Y_{e\mu} v_{3}e^{i\xi}}{d_{e}d_{\mu}} & -\dfrac{Y_{e\tau} v_{1}}{d_{e}d_{\tau}}\\
\dfrac{Y_{e\mu} v_{3}e^{i\xi}}{d_{e}d_{\mu}} & 0 & 0\\
-\dfrac{Y_{e\tau} v_{1}}{d_{e}d_{\tau}} & 0 & \dfrac{Y_{\tau \tau} v_{2}}{d_{\tau}^{2}}\\
\end{pmatrix},
\label{eqn14}
\end{eqnarray}
which corresponds to $D_{1}$ texture of $M_{\nu}^{-1}$ studied in Ref. \cite{Singh}.\\
Also, if the charge assignment of $\Phi_{2}$ under  $U(1)_{L_{\mu}-L_{\tau}}$ and $Z_{4}$ are replaced by $-2$ and $1$, respectively, then we obtain, $M_{R}$, given by
\begin{eqnarray}
M_{R} = 
\begin{pmatrix}
M & Y_{e\mu} v_{3}  & Y_{e\tau} v_{1} e^{i\xi}\\
Y_{e\mu} v_{3} & Y_{\mu \mu} v_{2}  & 0 \\
Y_{e\tau} v_{1}e^{i\xi} & 0  & 0\\
\end{pmatrix},
\label{eqn15}
\end{eqnarray}
while the mass matrices $M_{\ell}$ and $M_{D}$ remains diagonal. The two-zero texture of $M_{\nu}^{-1}$ obtained using Eqn.(\ref{eqn15}) corresponds to $E_{1}$ texture with zeroes at ($2,3$) and ($3,3$) place \cite{Singh}.

\section{Numerical Analysis and Discussion}

\noindent In this section, as a representative case, we perform the numerical analysis of texture $D_1$ obtained in Eqn.(\ref{eqn14}) in light of muon ($g-2$) and neutrino oscillation data (Table \ref{tabx}). The $D_1$ texture defined in Eqn.(\ref{eqn14}) corresponds to two-zero texture $M_{\nu}$ with zeroes at ($1,1$) and ($1,3$). $M_{\nu}$ is numerically diagonalised by a unitary matrix $U$ such that $UM_{\nu}U^{T}=\text{diag}(m_{1},m_{2},m
_{3})$ and the neutrino mixing angles can be obtained using
\begin{eqnarray}
\sin^{2}{\theta_{13}}=\left|U_{13}\right|^{2}, \hspace{0.5cm} \sin^{2}{\theta_{23}}=\dfrac{\left|U_{23}\right|^{2}}{1-\left|U_{13}\right|^{2}}, \hspace{0.5cm} \sin^{2}{\theta_{12}}=\dfrac{\left|U_{12}\right|^{2}}{1-\left|U_{13}\right|^{2}}.
\label{eqn16}
\end{eqnarray}

Also, the amount of $CP$ violation manifested in Jarlskog invariant ($J_{CP}$)\cite{Jarlskog1,Jarlskog2} is defined as
\begin{eqnarray}
J_{CP}=\text{Im}\left[U_{11}U_{22}U^{*}_{12}U^{*}_{21}\right]=s_{23}c_{23}s_{12}c_{12}s_{13}c^{2}_{13}\sin{\delta},
\label{eqn17}
\end{eqnarray}
while other two rephasing invariants $I_{1}$ and $I_{2}$ are given by
\begin{eqnarray}
I_{1}=\text{Im}\left[U^{*}_{11}U_{12}\right]=c_{12}s_{12}c^{2}_{13}\sin\left({\dfrac{\alpha_1}{2}}\right), \hspace{0.5cm} I_{2}=\text{Im}\left[U^{*}_{11}U_{13}\right]=c_{12}s_{13}c_{13}\sin\left(\dfrac{\alpha_2}{2}-\delta\right),
\label{eqn18}
\end{eqnarray}

\begin{figure}[t]
    \centering
   \begin{tikzpicture}
\begin{feynman}
\vertex at (3,0) (i1) {\(\mu\)};
\vertex at (-3,0) (i2) {\(\mu\)};
\vertex at (0, 3) (a);
\vertex at (2.3,0.7) (b);
\vertex at (-2.3,0.7) (c);
\vertex at (0,0.4) () {\(Z_{\mu\tau}\)};
\vertex at (0,5.5) (d) {\(\gamma\)};
\diagram*{
(i1) -- [fermion] (a) -- [fermion] (i2),
(b) -- [photon] (c),
(a) -- [photon] (d),
};
\end{feynman}
\end{tikzpicture}
    \caption{One loop Feynman diagram mediated by extra gauge boson $Z_{\mu\tau}$ contributing to muon $(g-2)$.}
    \label{fig:g2}
\end{figure}
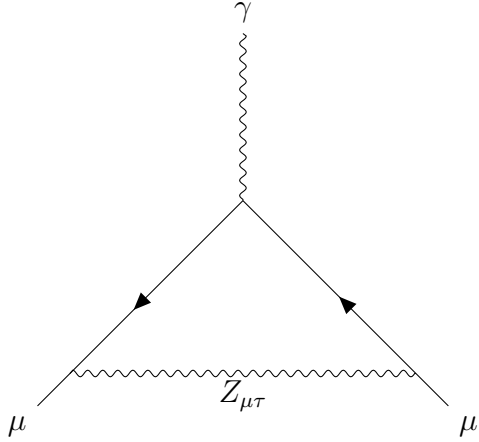

\noindent where $\alpha_1,\alpha_2$ are Majorana phases. Furthermore, the gyromagnetic ratio ($g$-factor) of the muon is the quantity which relates its spin ($\Vec{s}$) to its magnetic moment ($\vec{\mu}$) as given by
\begin{eqnarray}
\vec{\mu}=g\left(\dfrac{q}{2m_{\mu}}\right)\Vec{s},
\label{eqn21}
\end{eqnarray}
where, $q$ is muon charge and $m_{\mu}$ is muon mass. In Dirac's theory of charged spin-half particles, the gyromagnetic ratio is $g=2$. However, the recent developments at FermiLab hint towards non-trivial interactions of muon with BSM fields.
The higher-order radiative corrections can generate additional contributions to magnetic moment of muon parameterised as 
\begin{eqnarray}
g=2_{\text{Dirac}}(1+a_{\mu})\hspace{1cm} \text{and} \hspace{1cm}
a_{\mu}=\dfrac{1}{2}\left(g-2\right).
\label{eqn22}
\end{eqnarray}

\noindent The correction $a_{\mu}$ to the Dirac's predictions is called the anomalous magnetic moment. With in SM, the contribution to the anomalous magnetic moment of muon may comes from: (a) quantum electrodynamic $(\textit{QED})$ contributions (b) electroweak $(\textit{EW})$ contributions (c) hadronic vacuum polarisation contributions (d) hadronic light-by-light scattering contributions. As explained earlier, the SM is not consistent with the recent results on muon $(g-2)$ at FermiLab. Therefore, beyond standard model contribution is required to explain muon anomalous magnetic moment. In this model, the additional contribution to muon magnetic moment arises at one-loop (Fig.(\ref{fig:g2})) mediated by $U(1)_{L_{\mu}-L_{\tau}}$ gauge boson $Z_{\mu\tau}$ and is given  by\cite{AnomalousFormula1,AnomalousFormula2}
\begin{eqnarray}
\Delta a_{\mu}=\dfrac{\alpha'}{2\pi}\int_{0}^{1} dx\dfrac{2m^{2}_{\mu}x^{2}(1-x)}{x^{2}m^{2}_{\mu}+(1-x)M_{Z_{\mu\tau}}},
\label{g-2}
\end{eqnarray}

\begin{figure}
\centering
\begin{subfigure}{.5\textwidth}
  \centering
  \includegraphics[width=0.9\linewidth]{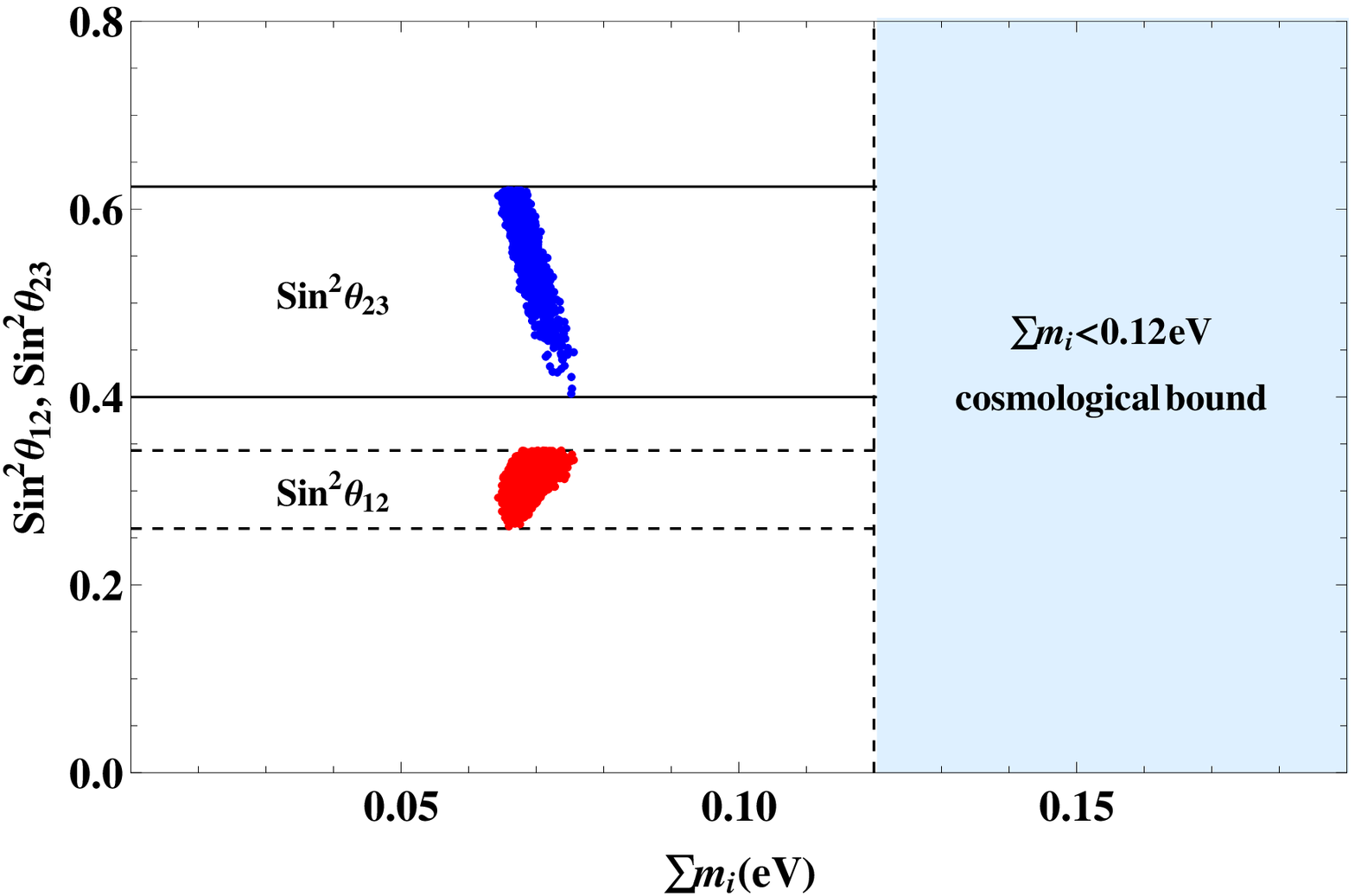}
  \subcaption*{2(a)}
  \label{fig:sub1}
\end{subfigure}%
\begin{subfigure}{.5\textwidth}
  \centering
  \includegraphics[width=0.9\linewidth]{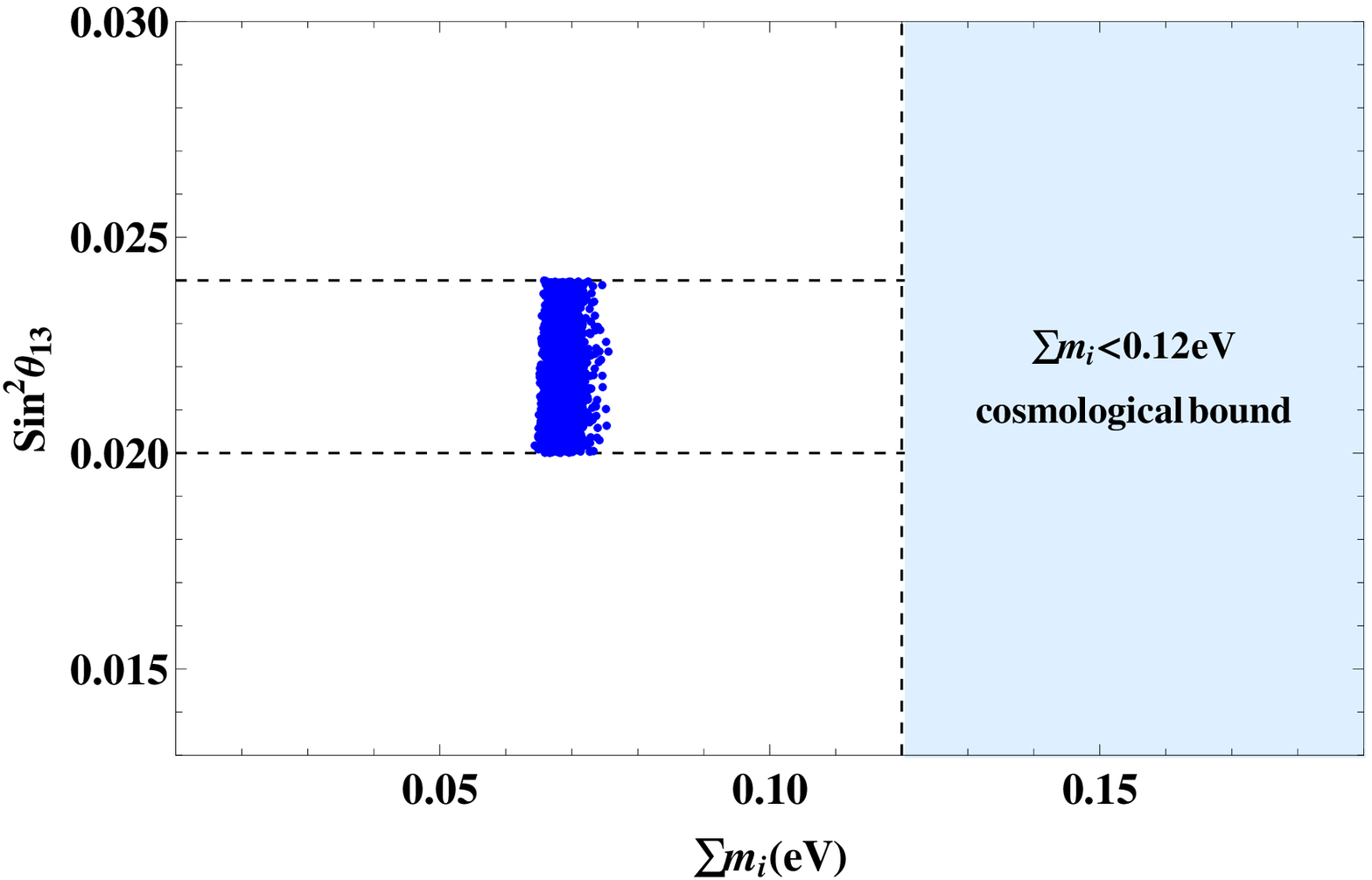}
  \subcaption*{2(b)}
  \label{fig:sub2}
\end{subfigure}
\caption{The model predictions of the neutrino mixing angles as correlations plots between sum of neutrino masses ($\sum m_i$) and mixing angles. The horizontal lines are 3$\sigma$ experimental bounds on the respective mixing angle. The shaded region is excluded by the cosmological bound on sum of neutrino masses.}
\label{fig2}
\end{figure}

\noindent where $\alpha'=\dfrac{g^{2}_{\mu\tau}}{4\pi}$ is the structure constant and $m_{\mu}$ is the mass of muon. The new gauge boson, $Z_{\mu\tau}$, gets mass after scalar singlet fields $(\Phi_{i})$ acquires $vevs$. Also, non-trivial neutrino mixing matrix is induced by $vevs$ of scalar singlet fields (through $M_{R}$), thus, connecting explanation of muon $(g-2)$ to neutrino phenomenology.

\begin{table}[t]
    \centering
    \begin{tabular}{|l| l l| l l|}
    \hline
  Parameter &$\pm 1\sigma$ range (NH)&$\pm 1\sigma$ range (IH)& $3\sigma$ range (NH) & $3\sigma$ range (IH)\\
\hline  
 $\sin^{2}{\theta_{12}}$&$0.304^{+0.013}_{-0.012}$&$0.304^{+0.012}_{-0.012}$ & 0.269-0.343 & 0.269-0.343 \\
\hline
 
$\sin^{2}{\theta_{13}^{2}}$&$0.02220^{+0.00068}_{-0.00062}$&$0.02238^{+0.00064}_{-0.00062}$& 0.02060-0.02435& 0.02053-0.02434\\
\hline
    
$\sin^{2}{\theta_{23}^{2}}$&$0.573^{+0.018}_{-0.023}$&$0.578^{+0.017}_{-0.021}$& 0.405-0.624& 0.410-0.623\\
\hline
   
$\frac{\Delta m_{12}^{2}}{10^{-5}\text{eV}^{2}}$ & $7.42^{+0.21}_{-0.20}$&$7.42^{+0.21}_{-0.20}$& 6.82-8.04  & 6.62-8.04\\
\hline
   
$\frac{\Delta m_{23}^{2}}{10^{-3}\text{eV}^{2}}$&$2.515^{+0.028}_{-0.028}$&$2.498^{+0.028}_{-0.029}$& 2.431-2.598 &-2.584- -2.413\\
\hline

\end{tabular}
    \caption{ The neutrino oscillation data from global fit used in the numerical analysis\cite{neutrinodata}.}
    \label{tabx}
\end{table}

\noindent There are eight free parameters in inverse neutrino mass matrix ($M_{\nu}^{-1}$), given as, $M_{ee}$, $V_{1}$=$Y_{e\mu}v_{1}$, $V_{2}=Y_{e\tau}v_{2}$, $V_{3}=Y_{\tau\tau}v_{3}$, $d_{e}$, $d_{\mu}$, $d_{\tau}$ and $\xi$. In order to obtain the predictions on neutrino oscillation parameters and muon $(g-2)$ anomaly for $D_{1}$ texture, we have randomly varied all the free parameters with uniform distribution in the ranges

\begin{equation}
\left.  
\begin{split}
d_{e}, d_{\mu}, d_{\tau}=\left( 10^{-5}-10^{-3}\right) \text{GeV}, \\
V_{1},  V_{2},  V_{3} = \left(1-280\right) \text{GeV}, \\
M= \left(1-10^{4}\right) \text{GeV}, \\
\xi= \left(0-360\right)^{\circ}.\\
\end{split}
\right\}  
\label{eqn20}
\end{equation} 
We have numerically diagonalized  $M_{\nu}$ to obtain the neutrino mixing matrix $U$. The predictions for neutrino mixing angles obtained from Eqn.(\ref{eqn16}) are compared with $3\sigma$ ranges given in Table \ref{tabx} to ascertain the allowed parameter space of the model. \\

\begin{figure}[t]
    \centering
    \includegraphics[width=7.5cm]{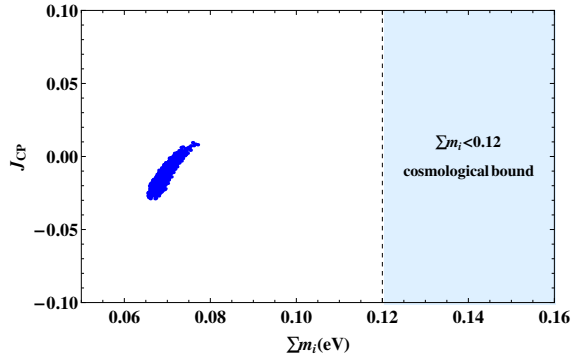}
    \caption{The correlation plot between $\left(\sum m_{i}-J_{CP}\right)$.  The shaded region is excluded by the cosmological bound on sum of neutrino masses.}
    \label{fig3}
\end{figure}

\begin{figure}
\centering
\begin{subfigure}{.5\textwidth}
  \centering
  \includegraphics[width=0.9\linewidth]{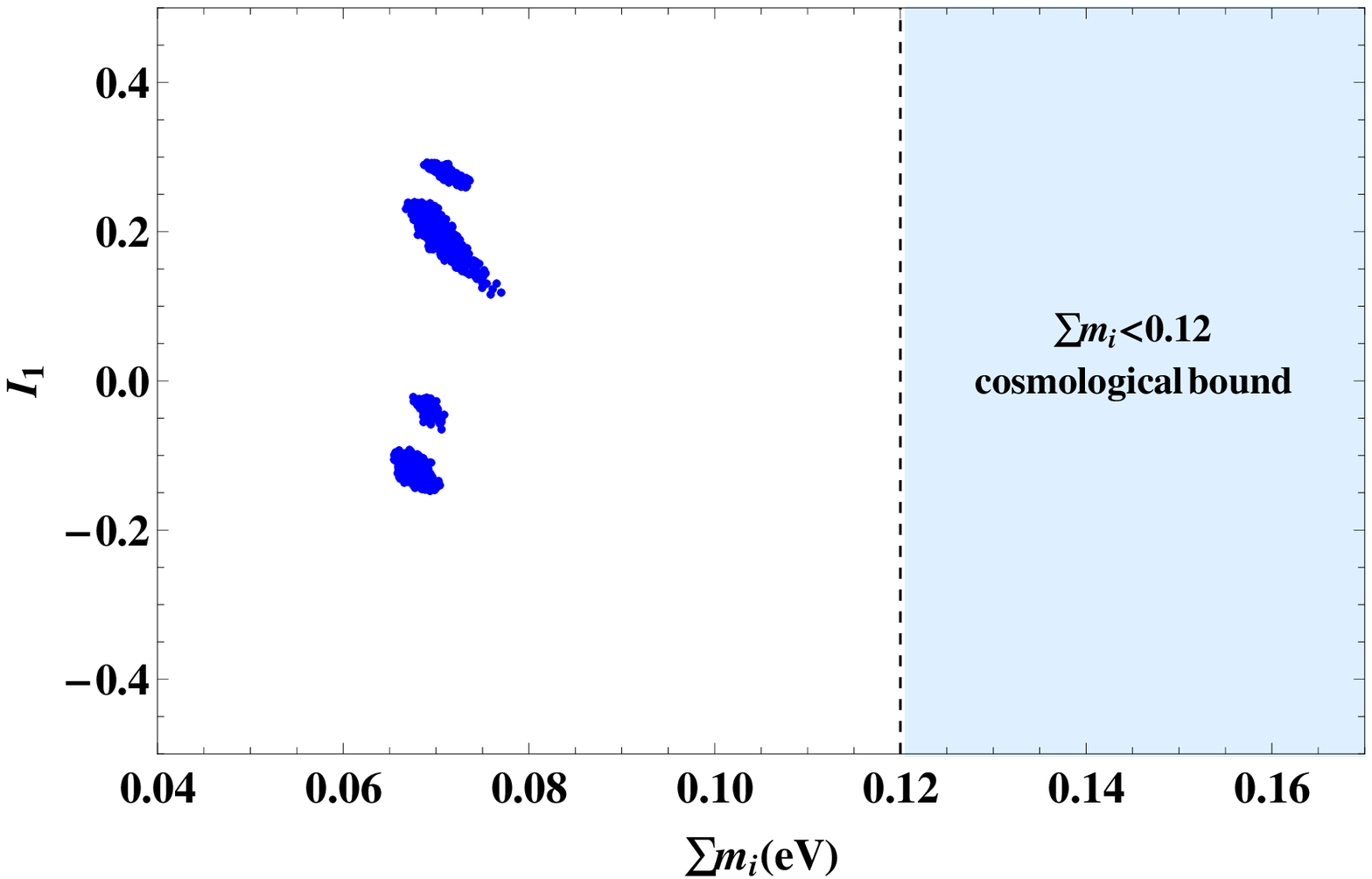}
  \subcaption*{4(a)}
  \label{fig:sub1}
\end{subfigure}%
\begin{subfigure}{.5\textwidth}
  \centering
  \includegraphics[width=0.9\linewidth]{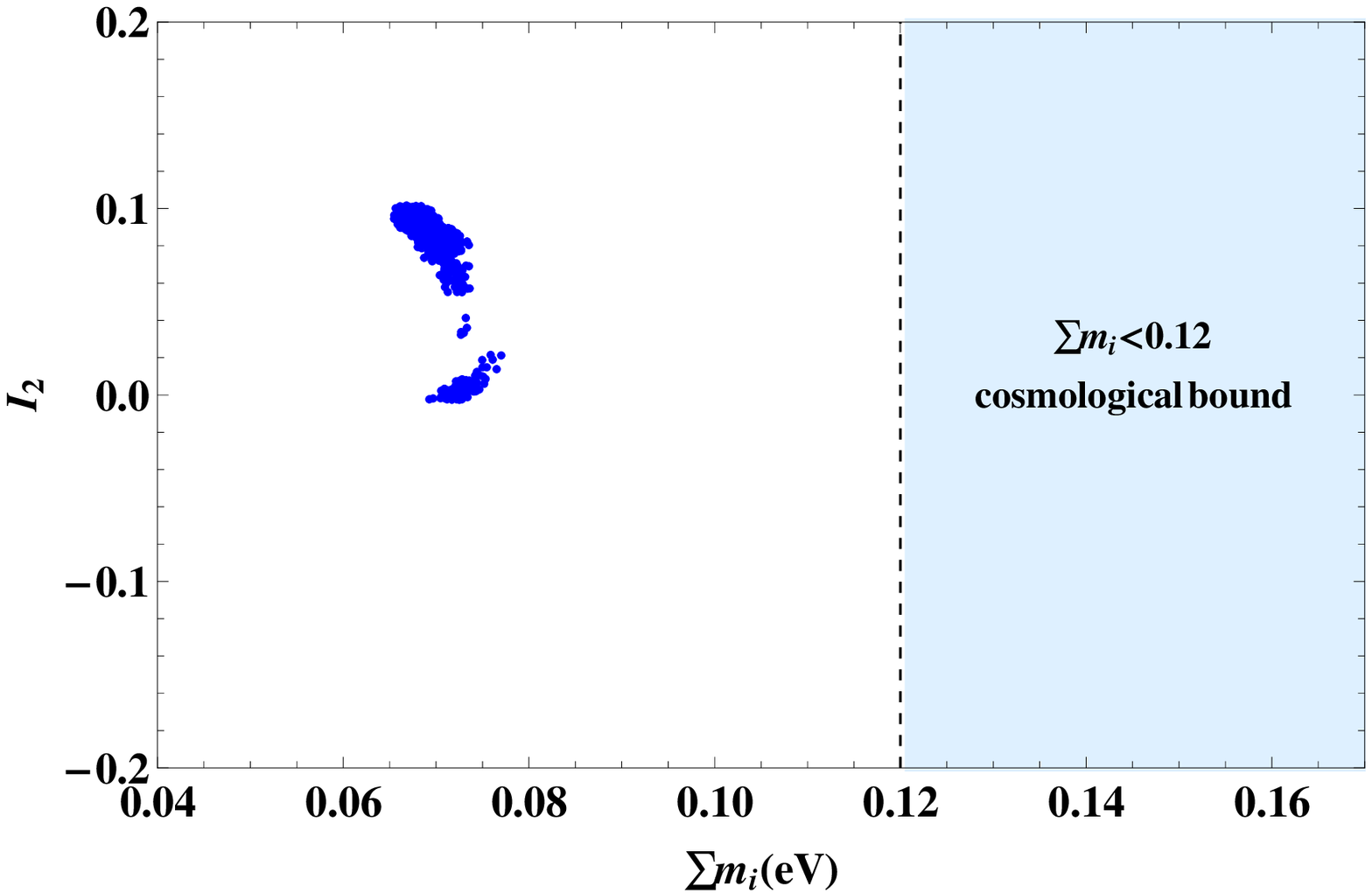}
  \subcaption*{4(b)}
  \label{fig:sub2}
\end{subfigure}
\caption{The correlation plots between $\sum m_{i}$ and  $CP$ invariants $I_1$(4(a)), $I_2$ (4(b)). The shaded region is excluded by the cosmological bound on sum of neutrino masses.}
\label{fig4}
\end{figure}

\noindent In Fig.\ref{fig2}(a), we have depicted the correlation between sum of neutrino masses $\sum m_{i}$ versus $\sin^{2}{\theta_{12}}$ and $\sin^{2}{\theta_{23}}$ while Fig.\ref{fig2}(b) shows the correlation between $\sum m_{i}$ versus $\sin^{2}{\theta_{13}}$ at 3$\sigma$. It is evident that the model is consistent with the neutrino oscillation data on the mixing angles and predicts sum of neutrino masses $\sum m_i$ to be with in the range $0.065\lesssim \sum m_{i}(\text{eV}) \lesssim 0.075$. In Fig.\ref{fig3}, we have given the correlation plot of $\left(\sum m_{i}-J_{CP}\right)$. $J_{CP}$ lies in the range $-0.03\leq J_{CP} \leq 0.01$. The predictions for other two $CP$ rephasing invariants $I_1$ and $I_{2}$ are shown in Fig.\ref{fig4}(a) and (b), respectively. It is evident from Fig.\ref{fig4}(a) that $I_1=0$ is disallowed implying $D_1$ texture is necessarily $CP$ violating.

\begin{figure}[t]
    \centering
    \includegraphics[width=10cm]{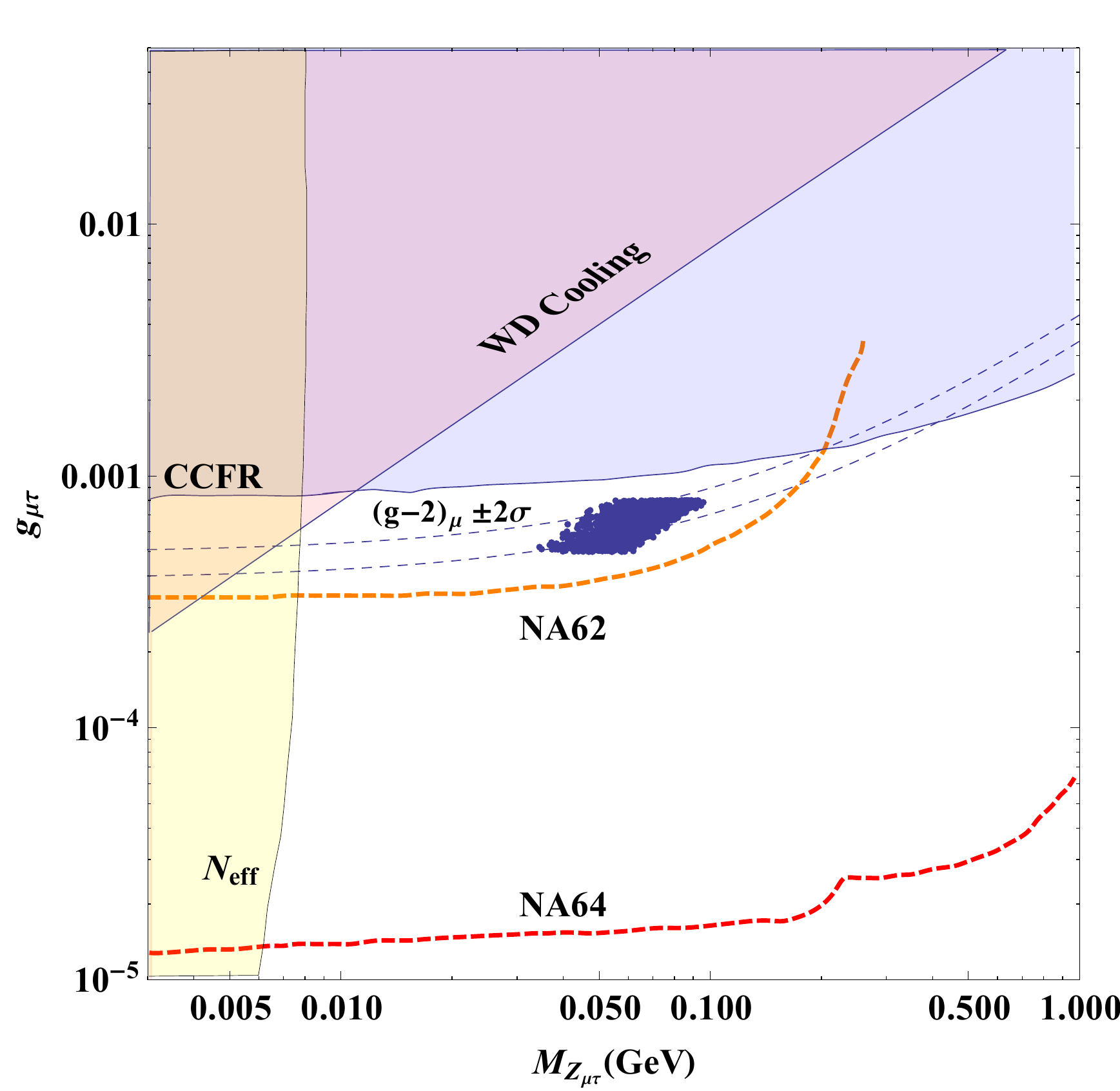}
    \caption{The allowed parameter space of the model in ($M_{Z_{\mu\tau}}-g_{\mu\tau}$) plane accommodating muon ($g-2$) and neutrino oscillation data. The exclusion regions from various experiments are, also, shown.}
    \label{fig5}
\end{figure}

Using the Eqn.(\ref{g-2}), we calculate the $Z_{\mu\tau}$ contribution to $\Delta a_{\mu}$ which has been shown in $M_{Z_{\mu\tau}}-g_{\mu\tau}$ plane in Fig.\ref{fig5}. The gauge coupling is randomly varied in the range $10^{-4}-10^{-3}$. It is evident from Fig.\ref{fig5} that the model accommodates the observed muon ($g-2$) for $M_{Z_{\mu\tau}}$ in the range ($0.035$ GeV-$0.100$ GeV) and $g_{\mu\tau}\approx\mathcal{O}(10^{-4}$), which is consistent with constraints coming from experiments like COHERENT\cite{COHERENT1,COHERENT2}, BABAR\cite{BaBar} and CCFR\cite{CCFR}. The sensitivities of future experiments NA62\cite{NA62} and NA64\cite{NA64A,NA64B} are, also, shown in Fig.\ref{fig5}. The upper left triangular region is excluded by the astrophysical bound from cooling of white dwarf (WD)\cite{WD}.

\textbf{Benchmark point:}
For the input parameters
\begin{eqnarray}
\nonumber
&&\left(d_{e},d_{\mu},d_{\tau}\right)\cross10^{-5}=(2.38,1.57,2.19) \text{GeV},\\
\nonumber
&&\left(V_{1},V_{2},V_{3},M\right)=(65.6,15.0,37.1,28.1) \text{GeV},\\
\nonumber
&&\xi=309.54^{\circ},\\ \nonumber
&& g_{\mu\tau}=5.2\cross10^{-4},
\end{eqnarray}

the corresponding values of mass-squared differences, mixing angles, $M_{Z_{\mu\tau}}$ and $\Delta a_\mu$ are

\begin{eqnarray}
\nonumber 
&&\Delta m^{2}_{23}=2.45\cross 10^{-3}\text{eV}^2; \hspace{0.1cm}\Delta m^{2}_{12}=7.53\cross 10^{-5}\text{eV}^2,\\
\nonumber
&&\text{sin}^{2}{\theta_{13}}=0.022;\hspace{0.1cm}\sin^{2}{\theta_{12}}=0.32;\hspace{0.1cm}\sin^{2}{\theta_{23}}=0.58,\\
&&\nonumber \hspace{0.1cm}M_{Z_{\mu\tau}}=42.38 \text{MeV};\Delta a_{\mu}=3.42\cross10^{-11}.
\end{eqnarray} 

\section{Conclusions}
In this work, we have realised two-zero textures of $M_{\nu}^{-1}$ with anomaly free gauged $U(1)_{L_{\mu}-L_{\tau}}$ extension of SM in light of the muon ($g-2$) anomaly. We have extended the SM field content by adding three scalar singlets ($\Phi_{i}$) and three right-handed neutrinos $(N_{e},N_{\mu},N_{\tau})$. $U(1)_{L_{\mu}-L_{\tau}}$ symmetry is broken as the new scalar singlets acquire $vevs$, thereby, giving mass to new $U(1)_{L_{\mu}-L_{\tau}}$ gauge boson $Z_{\mu\tau}$. The two-zeros in $M_{R}$ corresponds to two-zeros in $M_{\nu}^{-1}$ in the diagonal charged lepton and Dirac mass basis. Also, the non-trivial neutrino mixing depends on the structure of $M_{R}$. Thus, the right-handed Majorana neutrino mass matrix connects the low energy neutrino phenomenology with $M_{Z_{\mu\tau}}$ contributing to muon anomalous magnetic moment. We have scanned the model parameter space and have found the model consistent with the neutrino oscillation data within $3\sigma$ ranges. The Jarlskog $CP$ rephasing invariant, $J_{CP}$, lies in the range $-0.03\leq J_{CP} \leq 0.01$. The texture is found to be necessarily $CP$ violating as $I_1=0$ is disallowed. The model predicts the mass of new gauge boson $(M_{Z_{\mu\tau}})$ in the range $0.035 \text{GeV}\leq M_{Z_{\mu\tau}} \leq 0.100 \text{GeV}$ for gauge coupling ($g_{\mu\tau}$) between $5\times 10^{-4}\leq g_{\mu\tau}\leq 8\times 10^{-4}$ which is consistent with constraints from experiments such as CCFR, COHERENT, BABAR, NA62 and NA64.

\section*{Acknowledgments}
 M. K. acknowledges the financial support provided by Department of Science and Technology(DST), Government of India vide Grant No. DST/INSPIRE Fellowship/2018/IF180327. The authors, also, acknowledge Department of Physics and Astronomical Science for providing necessary facility to carry out this work.

\newpage

\end{document}